\documentclass[11pt]{article}
\usepackage{extsizes}
\usepackage[super,sort&compress,comma]{natbib} 
\usepackage[version=3]{mhchem}
\usepackage[left=2.54cm, right=2.54cm, top=2.54cm, bottom=2.54cm]{geometry}
\usepackage{balance}
\usepackage{mathptmx}
\usepackage{sectsty}
\usepackage{graphicx} 
\usepackage{lastpage}
\usepackage[format=plain,justification=justified,singlelinecheck=false,font={stretch=1.125,small,sf},labelfont=bf,labelsep=space]{caption}
\usepackage{float}
\usepackage{fancyhdr}
\usepackage{fnpos}
\usepackage[english]{babel}
\addto{\captionsenglish}{%
  
}
\usepackage{array}
\usepackage{droidsans}
\usepackage{charter}
\usepackage[T1]{fontenc}
\usepackage[usenames,dvipsnames]{xcolor}
\usepackage{setspace}
\usepackage[compact]{titlesec}
\usepackage{hyperref}
\usepackage{amsmath,bm}

\usepackage{epstopdf}

\definecolor{cream}{RGB}{222,217,201}

\usepackage[switch]{lineno}
\modulolinenumbers[1]

\DeclareUnicodeCharacter{2212}{-}

\begin{document}

\pagestyle{fancy}
\thispagestyle{plain}
\fancypagestyle{plain}{
\renewcommand{\headrulewidth}{0pt}
}

\makeFNbottom
\makeatletter
\renewcommand\LARGE{\@setfontsize\LARGE{15pt}{17}}
\renewcommand\Large{\@setfontsize\Large{12pt}{14}}
\renewcommand\large{\@setfontsize\large{10pt}{12}}
\renewcommand\footnotesize{\@setfontsize\footnotesize{7pt}{10}}
\makeatother

\renewcommand{\thefootnote}{\fnsymbol{footnote}}
\renewcommand\footnoterule{\vspace*{1pt}%
\color{cream}\hrule width 3.5in height 0.4pt \color{black}\vspace*{5pt}} 
\setcounter{secnumdepth}{5}

\makeatletter 
\renewcommand\@biblabel[1]{#1}            
\renewcommand\@makefntext[1]%
{\noindent\makebox[0pt][r]{\@thefnmark\,}#1}
\makeatother 
\renewcommand{\figurename}{\small{Fig.}~}
\sectionfont{\sffamily\Large}
\subsectionfont{\normalsize}
\subsubsectionfont{\bf}
\setstretch{1.125} 
\setlength{\skip\footins}{0.8cm}
\setlength{\footnotesep}{0.25cm}
\setlength{\jot}{10pt}
\titlespacing*{\section}{0pt}{4pt}{4pt}
\titlespacing*{\subsection}{0pt}{15pt}{1pt}


\fancyfoot[RO]{\footnotesize{\sffamily{1--\pageref{LastPage} ~\textbar  \hspace{2pt}\thepage}}}
\fancyfoot[LE]{\footnotesize{\sffamily{\thepage~\textbar\hspace{3.45cm} 1--\pageref{LastPage}}}}
\fancyhead{}
\renewcommand{\headrulewidth}{0pt} 
\renewcommand{\footrulewidth}{0pt}
\setlength{\arrayrulewidth}{1pt}
\setlength{\columnsep}{6.5mm}
\setlength\bibsep{1pt}

\makeatletter 
\newlength{\figrulesep} 
\setlength{\figrulesep}{0.5\textfloatsep} 

\newcommand{\topfigrule}{\vspace*{-1pt}%
\noindent{\color{cream}\rule[-\figrulesep]{\columnwidth}{1.5pt}} }

\newcommand{\botfigrule}{\vspace*{-2pt}%
\noindent{\color{cream}\rule[\figrulesep]{\columnwidth}{1.5pt}} }

\newcommand{\dblfigrule}{\vspace*{-1pt}%
\noindent{\color{cream}\rule[-\figrulesep]{\textwidth}{1.5pt}} }

\makeatother


\vspace{1em}

\noindent\LARGE{\textbf{Quantifying Multiphase Flow of Aqueous Acid and Gas CO2 in Deforming Porous Media Subject to Dissolution$^\dag$}} \\

\noindent\large{Rafid Musabbir Rahman, Carson Kocmick, Colin Shaw and Yaofa Li$^{\ast}$} \\

\noindent\normalsize{Mineral dissolution in porous media coupled with single- or multi-phase flows is pervasive in natural and engineering systems including carbon capture and sequestration (CCS) and acid stimulation in reservoir engineering. Dissolution of minerals occurs as chemicals in the solid phase is transformed into ions in the aqueous phase, effectively modifying the physical, hydrological and geochemical properties of the solid matrix, and thus leading to the strong coupling between local dissolution rate and pore-scale flow. However, our fundamental understanding of this coupling effect at the pore level is still limited. In this work, mineral dissolution is studied using novel calcite-based porous micromodels under single- and multiphase conditions, with a focus on the interactions of mineral dissolution with pore flow. The microfluidic devices used in the experiments were fabricated in calcite using photolithography and wet etching. These surrogate porous media offer precise control over the structures and chemical properties and facilitate unobstructed and unaberrated optical access to the pore flow with µPIV methods. The preliminary results provide a unique view of the flow dynamics during mineral dissolution. Based on these pore-scale measurements, correlations between pore-scale flow and dissolution rates can be developed for several representative conditions.} 


\renewcommand*\rmdefault{bch}\normalfont\upshape
\rmfamily
\section*{}
\vspace{-1cm}


\footnotetext{Mechanical \& Industrial Engineering Department, Montana State University, Bozeman, Montana, USA. Fax: 406-994-6292; Tel: 406-994-6773; E-mail: yaofa.li@montana.edu}
\footnotetext{{$\ast$} Corresponding author.}








\vspace{1cm}
\section{Introduction}

Reactive dissolution of minerals in porous media coupled with single- or multi-phase flows is pervasive in natural and engineering systems \cite{Lund1975, Daccord1987, Drever1994, Adamo2002, Spycher2003, White2003, Maher2004, Maher2006, Molins2012, Nogues2013, Ott2015, Liu2015, Snippe2017, Gray2018, Soulaine2018, Voltolini2019}. In subsurface environments, the solid porous matrix is composed of various types of minerals, through which subsurface water flows. Dissolution of minerals occurs as chemicals in the solid phase is transformed into ions in the aqueous phase, effectively modifying the physical, hydrological and geochemical properties of the solid matrix as well as the chemistry in the aqueous phase. These processes play a defining role in a broad range of applications including: ($i$)~carbon capture and sequestration (CCS) \cite{Ott2015, Voltolini2019}; ($ii$)~acid stimulation in reservoir engineering \cite{Lund1975, Daccord1987}; ($iii$)~soil formation and contaminant transport in soil \cite{Spycher2003, Adamo2002}; ($iv$)~development of underground cave systems \cite{Mylroie1990}; and ($v$)~global carbon and nutrient cycling \cite{Kump2000, Carey2005}. For instance, CCS is considered as a viable technology to reduce carbon emissions to the atmosphere, thus effectively mitigating global climate change. However, injection of CO$_2$ into geologic formations leads to dissolution of minerals comprising reservoirs rocks, potentially creating leakage pathways that threaten the safety and security of CO$_2$ storage \cite{Ott2015, Voltolini2019}. In agriculture, mineral dissolution supplies nutrients to soil \cite{Carey2005, Williams1997, Cosby1985, Schnoor1986}, but at the same time it can also cause the release of heavy metals leading to soil contamination \cite{Spycher2003, Adamo2002}. In reservoir stimulation, reactive fluids such as hydrochloric acid (HCl) are injected into damaged wellbores to increase its permeability and thus the well productivity \cite{Lund1975}. Therefore, a comprehensive understanding of mineral dissolution is crucial to successfully modeling, predicting, controlling and optimizing many those processes.

However, a full understanding of reactive dissolution in porous media is still lacking due to a number of complications. The porous materials are often of high \textit{heterogeneity} structurally and chemically, causing complex flow and reaction patterns. The ranges of the underlying spatial and temporal scales for practical problems are enormous, and processes at fine (pore) scales are known to meaningfully impact dissolution at much larger (reservoir) scales. In fact, one of the most quoted challenges in the literature is that dissolution rates in porous media measured in the lab are typically orders of magnitude higher than those observed in the field, often referred to as the ``lab–field discrepancy" \cite{Drever1994, White2003, Maher2006, Deng2018, Li2007, Li2008, Kim2011, Molins2012, Harrison2017, Menke2015, Swoboda1993, Menke2016}. The mismatch not only poses strong challenges in developing accurate predictive models based on the rate laws developed in laboratory, but also highlights a lack of fundamental understanding of mineral dissolution in porous media.

Dissolution of minerals in porous media is subject to chemical reaction, advection and diffusion. Field-based dissolution rates are usually estimated by mass balance approaches in soil profiles and watersheds, which are often subject to transport limitations \cite{White2003, Bricker2003}, whereas the laboratory reaction rates are often measured in \textit{well-mixed} systems using freshly crushed mineral suspensions, where mass transport limitations in the aqueous phase are \textit{intentionally eliminated} through forced mixing \cite{Chou1984}. As such, several factors have been proposed or identified to be collectively responsible for this discrepancy, which fall into two groups: extrinsic ones including the density of reactive sites and secondary mineralization, and intrinsic ones including heterogeneous pore flow and microbial processes \cite{Deng2018, Liu2015}. While the exact reason(s) and their relative contributions to the enormous discrepancy is still under debate, it is well accepted that that the laboratory measurement based on well-mixed systems (\textit{i.e., no concentration gradient}) does not appropriately represent the scenario in real fields, because: ($a$)~in fields, pore-level \textit{concentration gradient} can develop depending on the relative strengths of reaction and transport; ($b$)~accessibility of reactive surfaces depends strongly on the \textit{local flow} conditions. In fact, several studies have conjectured the lab-field discrepancy is primarily due to the \textit{concentration gradients} resulting from incomplete mixing within \textit{individual pores} that are in turn subject to \textit{heterogeneous flow fields at the microscopic scale} \cite{Deng2018, Li2007, Li2008, Molins2012, Menke2015, Menke2016}. Additionally, it has been shown that \textit{multiphase flow} is possible in mineral dissolution, which significantly affects the accessibility of reactive surfaces \cite{Soulaine2018, Song2018, Cheng2017, Bastami2016, Babaei2018, Mahmoodi2018, Snippe2017, Ott2015}. 

To further our understanding, quantifying pore flow is highly needed. Pore-scale processes in mineral dissolution are coupled through a non-linear feedback loop linking fluid flow, reactant transport, surface reaction and pore evolution (\textit{cf.} Fig. 2) \cite{Molins2012}. A positive feedback mechanism is well known to exist between local fluid flow and reactive dissolution, which is responsible for the creation of wormholes during acid stimulation \cite{Golfier2002, Lund1975, Hoefner1988}. Additionally, increasing evidences reinforce that \textit{pore flow} (\textit{i.e.}, advective transport) has very subtle effects on the development of concentration gradient and thus local dissolution rate \cite{Li2007, Li2008}. Pore flow is especially important when assessing and upscaling the pore-scale effects to a much larger domain, considering that for a given distance $l$ the required time would scale with $l$ for advective transport, but with $l^2$ for diffusive transport. 

Moreover, previous studies on reactive dissolution in porous media have been almost exclusively focused on single-phase flow conditions, although abundant evidences have shown that multiphase can play an important role \cite{Golfier2002, Bastami2016, Babaei2018, Mahmoodi2018, Snippe2017, Ott2015}. For instance, during acid stimulation of carbonate reservoirs under practical injection conditions, produced CO$_2$ can emerge as a separate phase, creating a multiphase environment for subsequent dissolution processes \cite{Soulaine2018, Song2018, Cheng2017}. An immiscible phase such as oil or gas can even be present prior to acid injection \cite{Babaei2018, Shukla2006}. Multiphase flow introduces effects like capillarity and flow instability and its complex phase configuration can alter the abundance of reactive surface areas thus affecting dissolution rate. Accounting for multiphase flow effects through innovative pore flow measurement is strongly needed to reshape our understanding of mineral dissolution.

To this end, single- and multiphase flows in porous media subject to simultaneous mineral dissolution are investigated in novel calcite-based porous micromodels. The microfluidic devices used in the experiments were fabricated in calcite using photolithography and wet etching. These surrogate porous media offer precise control over the structures and chemical properties and facilitate unobstructed and unaberrated optical access to the pore flow with µPIV methods. White field microscopy and the fluorescent micro-PIV method were simultaneously employed by seeding the water phase with fluorescent particles, in order to achieve simultaneous measurement of the velocity fields of the aqueous and structure evolution of the solid phase. Specifically, the objective of the study is to: 1)~design and fabricate calcite-based micromodels using microfabrication techniques; 2)~evaluate how local dissolution rate is affect by local flow field and multiphase flow.

\begin{figure*}[t]
\centering
\includegraphics[width=0.9\linewidth]{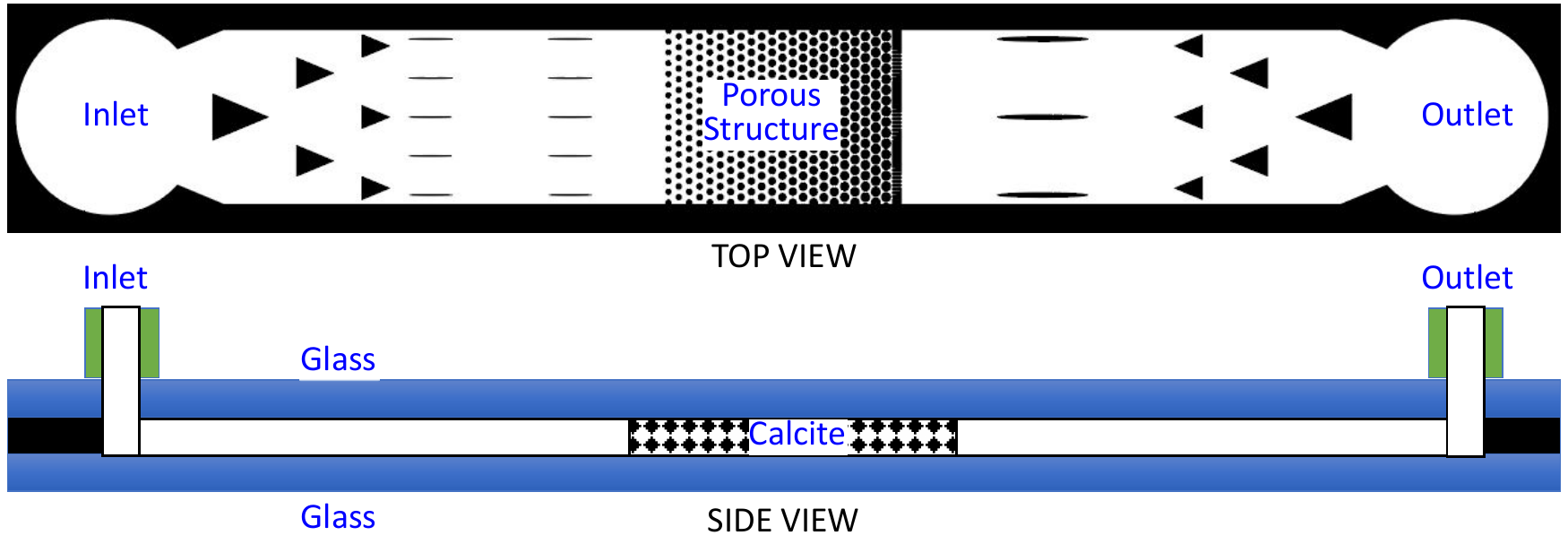}
\caption{Schematic diagram illustrating the design of the micromodel for the top [top] and side [bottom] views. The micromodel consists of 2 layers of glass with a thin calcite layer sandwiched in between, and the porous section is formed by regularly arranged circular posts.}
\label{fig:MicromodelDesign}
\end{figure*}

\section{Experimental Methods}
\subsection{Micromodel}
The micromodel consist of an inlet, an outlet, and a porous section, as shown in Fig.~\ref{fig:MicromodelDesign} [top]. While the inlet and outlet serve for fluid delivery, the porous section is the region of interest for flow and dissolution quantification. To mimic a porous media, the porous section is formed by regularly arranged circular posts with an overall size of 3.2\,mm $\times$ 4.8\,mm, and an porosity of approximately 45\%. From the side view (Fig.~\ref{fig:MicromodelDesign} [bottom]), the micromodel is constructed with three layers, with two glass layers sandwiching a calcite layer in between. The glass layers provide a structural support for the calcite layer, and also facilitate unobstructed and unaberrated optical access to the pore flow within the porous structure. The calcite layer presents an opportunity for us to effectively simulate the geochemistry and pore-scale fluid–rock interactions in real time.  

\begin{figure}
\centering
\includegraphics[width=0.9\linewidth]{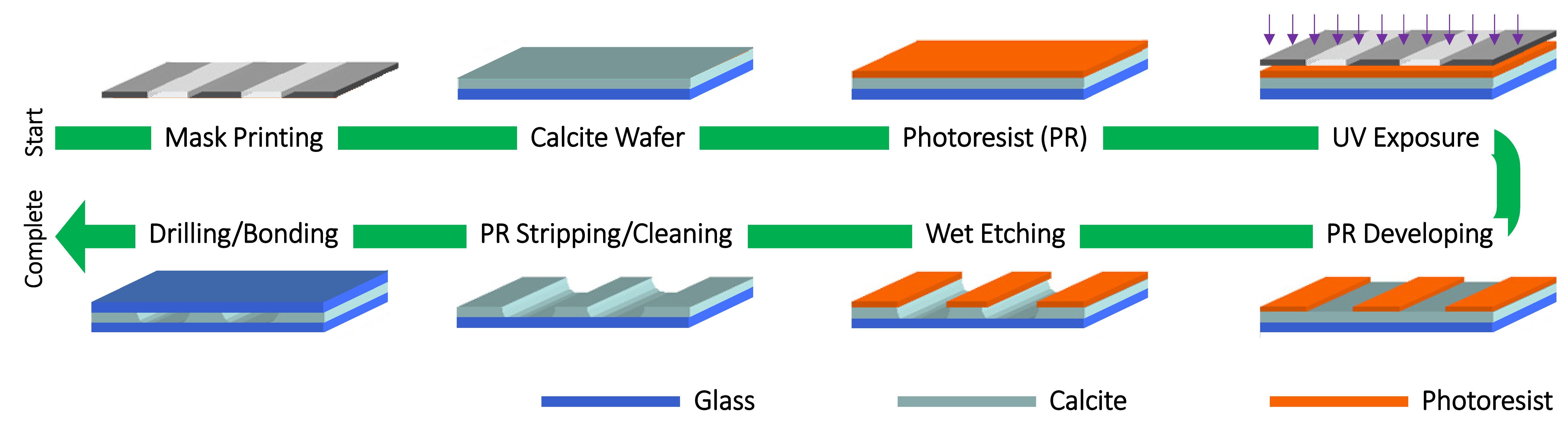}
\caption{Flowchart illustrating the major steps required to fabricate a calcite-based micromodel. A bulk calcite crystal was sectioned, double side polished, and grounded to 100\,$\mu$m. By photolithography, the porous pattern was transferred to the calcite wafer, which was then wet-etched in 0.4\% HCl solution. Two through holes were drilled. The wafer was then be cleaned and adhesively bonded to another glass wafer on the top to form closed flow channels.}
\label{fig:MicromodelFabrication}
\end{figure}

The fabrication procedures are illustrated in Fig.~\ref{fig:MicromodelFabrication}. First the calcite crystal was cleaned with acetone and attached to a reference glass using an epoxy glue. The calcite crystal was then sectioned using a thin-sectioning saw (Pelcon Automatic Thin Section Machine). The top surface was ground with three steel grinding rollers, which were coated with nickel bonded diamonds. The grinding process begins with the coarse roller embedded with 64\,$mu$m diamonds, continues on the intermediate roller with 32\,$mu$m diamonds and ends on the fine roller with 16\,$um$m diamonds. Following the grinding process, the top surface was attached to a clear microscopic slide using UV glue which was cured with UV light for 10 minutes while the sample was securely clamped. Using the same thin-sectioning machine, the reference glass was removed. The new open surface was then ground following the same procedures with the same 3 rollers. Finally, a polishing step was carried out using 6 $\mu$m grit for 2 minutes at 150\,rpm. The resulting thickness of the calcite sample is $\sim$31\,$\mu$m with an of uncertainty of 5.5\,$\mu$m. 

The calcite sample attached to the glass substrate was then ready for photolithography to transfer the designed pattern (Fig.~\ref{fig:MicromodelDesign}) to the sample. First, a spin coater (Laurell) was used to spin coat a 9\,$mu$m thick photoresist (AZ10XT) at 2000\,rpm for 45\,s. The coated calcite wafer was soft baked for 6\,minutes at 115\,$^{\circ}$C, following which the sample was allowed to rest for 1 hour to equilibrate with the environment. Second, the AB-M contact aligner was used to expose the photoresist at 15\,mW/cm$^2$ for 25\,s, which was then developed in AZ300 MIF developer for 15 minutes. The complete the photolithograpy, the sample was rinsed with DI water and dried with N2 stream. Finally, the porous structure was etched by immersing the sample in 0.4\% hydrochloric acid (HCl) for 8 minutes. To complete the etching process, the calcite was again rinsed with DI water and then dried with N2 gas (Fig.~\ref{fig:PreliminaryModel}). Following the etching process, a second microscope slide was bonded to form a closed micromodel. Two nanoports (IDEX Health \& Science) were attached to the micromodel for fluid delivery \cite{Li2019}

\begin{figure}[hbt]
\centering
\includegraphics[width=1\linewidth]{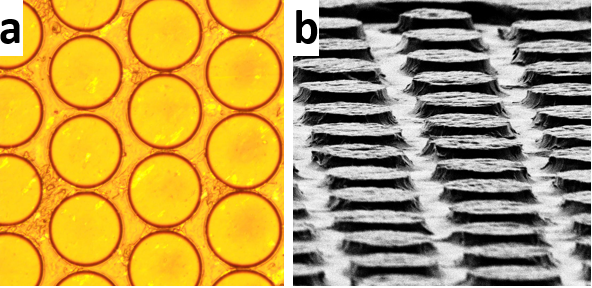}
\caption{(a)~A photo of the sample after photolithography illustrating the pattern; (b)~ An SEM photo of the resulting porous structure after etching.}
\label{fig:PreliminaryModel}
\end{figure}

\subsection{Experimental Procedure}
The working fluid used in this study was HCl with a concentration of 0.8\% to ensure an effective but controlled reaction with the solid calcite. To facilitate $\mu$PIV measurement in the aqueous phase, the HCl solution was seeded with fluorescent polystyrene particles of 1\,$mu$m in diameter (Invitrogen F8819), at a nominal volume fraction of $3 \times 10^{−4}$. The volume fraction of the solid particles is low so as to not appreciably change the aqueous phase properties and their density matched that of water so as to behave in a neutrally buoyant manner. Prior to each experiment, a new micromodel is cleaned and dried using a Nitrogen stream, and then the nanoports were connected to the plumbing tubings. The flow of the aqueous phase was then initiated, which was controlled by syringe pump (Harvard Apparatus PHD 2000).

\subsection{Image Acquisition and Processing}
In $\mu$PIV, two consecutive images of the particles are recorded with a prescribed time delay, ${\Delta}t$. Each image is then subdivided into small areas, called interrogation windows. The average displacement of the tracer particles $\bm{\Delta}x$ within each interrogation window is determined statistically from cross-correlation analysis of consecutive images. The velocity vector $\bm{u}$ for each interrogation window can then be obtained by dividing the displacement $\bm{\Delta}x$ with the prescribed ${\Delta}t$ \cite{Santiago1998}. To perform $\mu$PIV measurement, a green LED (Thorlabs Solis-525c) with a dominant wavelength of 525\,nm was used to excite the fluorescent particles in the aqueous phase. The fluorescence emitted from the tracer particles was passed through a $\lambda=575\pm13$\,nm bandpass filter and focused by an Olympus IX-71 microscope equipped with an objective lens with 4x magnification and 0.1 numerical aperture (NA), onto the detector of a high-speed scientific CMOS camera (Phantom VEO440). The camera was run at 100 frame per second, providing sufficient temporal resolution to capture the evolution of dynamic pore-scale phenomena. 

Owing to the multi-phase nature of some of the flows under study, the gas phase of the flow where no tracer particle is present needes to be masked out prior to image interrogation for quantification of the water velocity distribution \cite{Li2017, Li2021pore, Li2021particle}. This masking was performed based on a simple thresholding procedure complemented by manual adjustment when necessary \cite{Li2019}. For details regarding the image segmentation and image mask creation, the reader is referred to Refs.\cite{Li2019, Li2021particle}. To calculate the water velocity field, both frames in an acquired frame pair were first masked using the mask created in the previous steps and then interrogated in an in-house MATLAB code using a multi-pass, two-frame cross-correlation method \cite{Li2021particle}. Unless otherwise stated, the size of the interrogation windows was 32$\times$32 pixels, with 50\% overlap, which yielded a velocity vector spacing of 40\,$\mu$m and a spatial resolution of 80\,$\mu$m. 

\section{Preliminary Results}
While the micromodel fabrication and experiments are still undergoing, herein in this paper we present some preliminary results to showcase our study. It is worth noting that, the micromodel used in this preliminary study was fabricated in a slightly different way from what was described above. Specifically, to create the porous structures, instead of wet-etching, the micro-structure was micro-milled directly into a polished calcite crystal based on the patterns as shown in Fig.~\ref{fig:PreliminaryModel}. Nevertheless, the preliminary results provide a unique view of the flow dynamics during mineral dissolution.

\begin{figure}[hbt]
\centering
\includegraphics[width=0.7\linewidth]{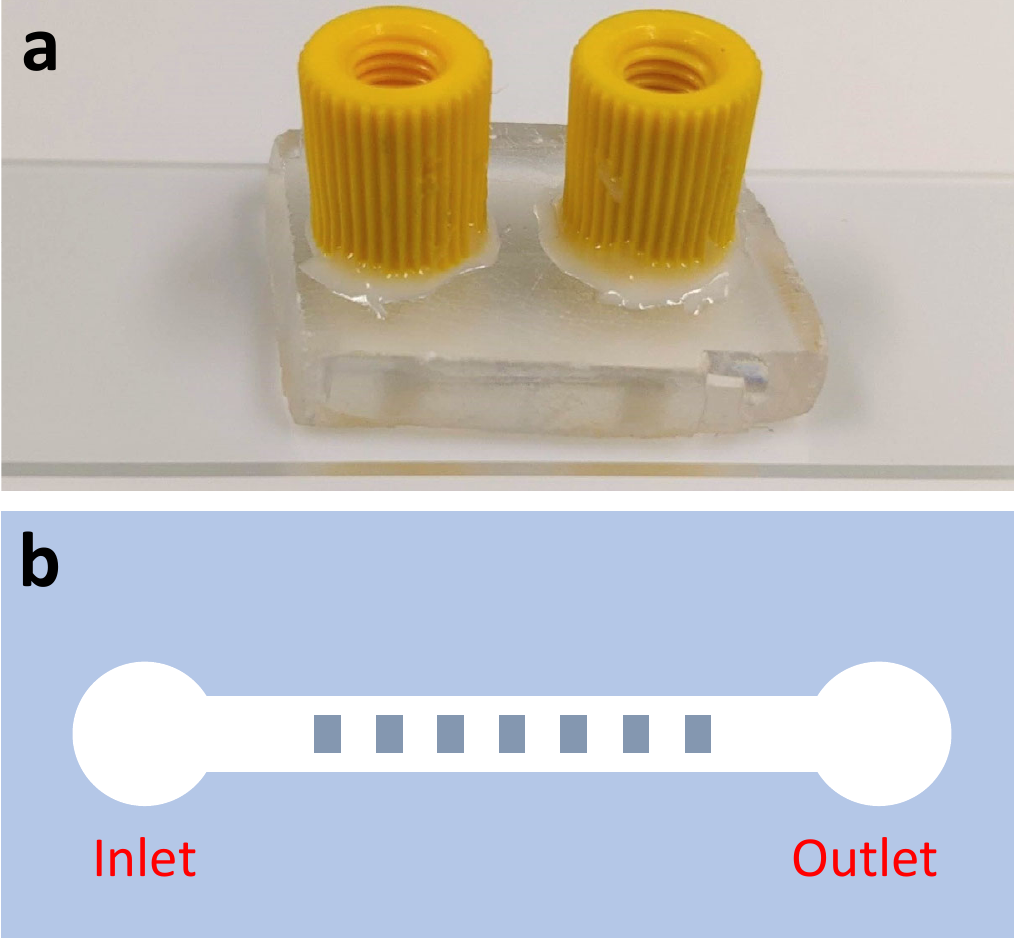}
\caption{(a)~A photo of the preliminary micromodel fabricated by micro-milling a piece of polished calcite crystal. The top layer is the polished calcite crystal, which is bonded to a glass slide, whereas the yellow parts are the nanoports used for fluid delivery. (b)~A schematic showing the machined porous structure in the preliminary micromodel.}
\label{fig:PreliminaryModel}
\end{figure}

\begin{figure}[hbt]
\centering
\includegraphics[width=0.9\linewidth]{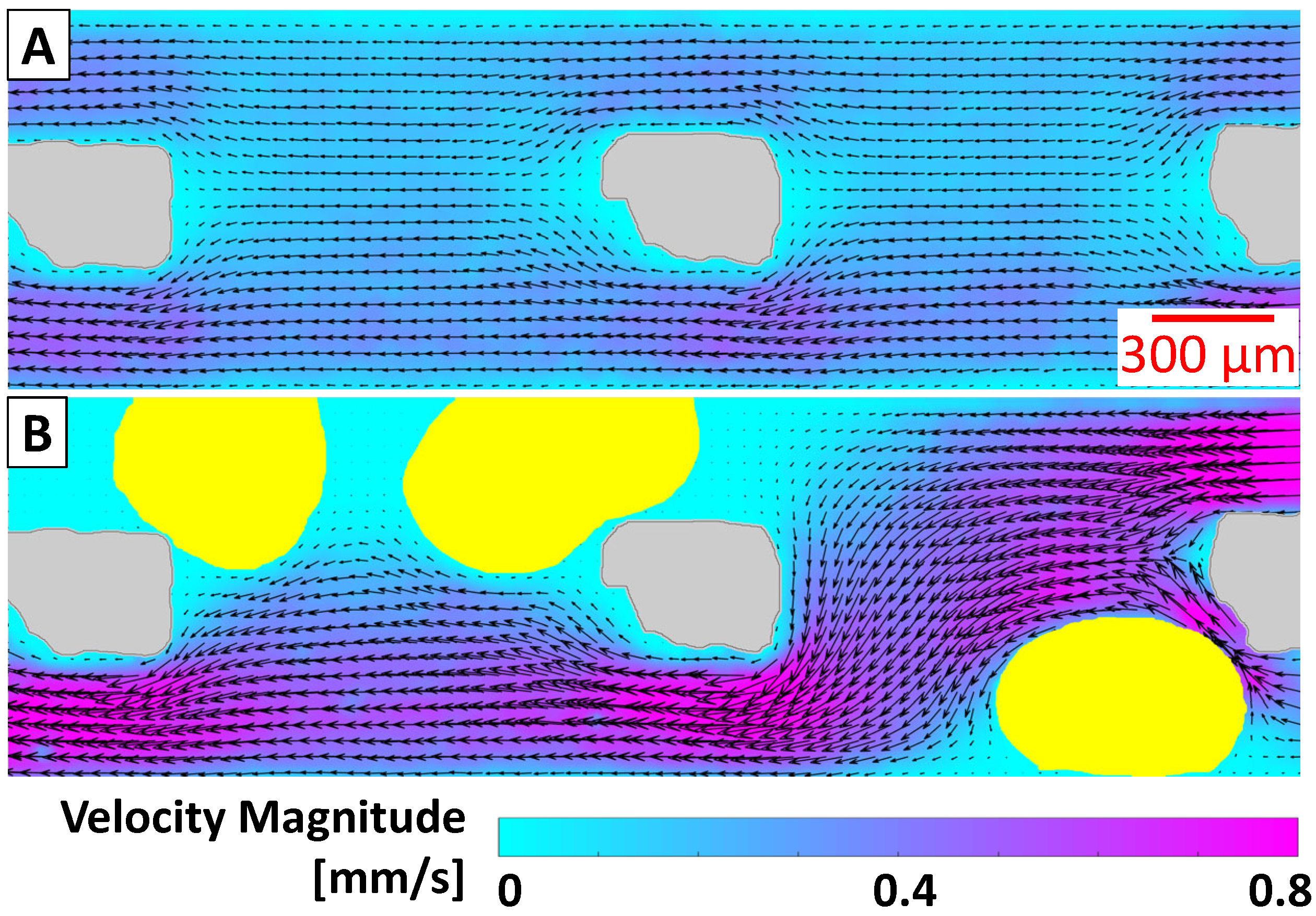}
\caption{Sample velocity fields of a single-phase flow of water (A) and a multiphase flow of water and air (B) through the calcite micromodel at the same flow rate. Contours and arrows indicate the velocity magnitude and direction, respectively, in the aqueous phase; and gray and yellow regions represent solid grains and gas bubbles, respectively.}
\label{fig:piv}
\end{figure}

Figures~\ref{fig:piv}a and \ref{fig:piv}b show the sample velocity fields of the pore-scale flow under single and multiphase flow conditions, respectively. When HCl concentration is low but the pore flow rate is relatively high, the reaction product (\textit{i.e.}, CO$_2$) is instantaneously dissolved in the aqueous phase leading to a single phase flow. However, when HCl concentration is high such that the produced CO$_2$ cannot be instantaneously dissolved, it will emerge as a separate phase, leading to a multiphase flow. While the flow field with single-phase flow is relatively simple, that flow is significantly modified in the multiphase flow case due to the presence CO$_2$ bubbles that are generated in-situ as a result of chemical reaction between the liquid and solid phases. The separate CO$_2$ phase is expected to not only divert the HCl flow, but also shied the solid surfaces from further reaction, thus significantly modifying the local dissolution pattern and rate.

The velocity fields generated by this technique also enables computations of various scalar fields, such as Péclet ($Pe$) and Damköhler number($Da$) fields, all involved in the mechanisms controlling the reactive transport and used as characterizating metrics of the relative importance of reaction, diffusion and advection. The Péclet number ($Pe$) and the Damköhler number are ($Da$) defined as \cite{Tartakovsky2008, Battiato2011}:

\begin{equation}\label{eqn:Pe}
Pe = Vl/D
\end{equation}

\begin{equation}\label{eqn:Da}
Da=kl/V
\end{equation}

\begin{figure}[hbt]
\centering
\includegraphics[width=0.9\linewidth]{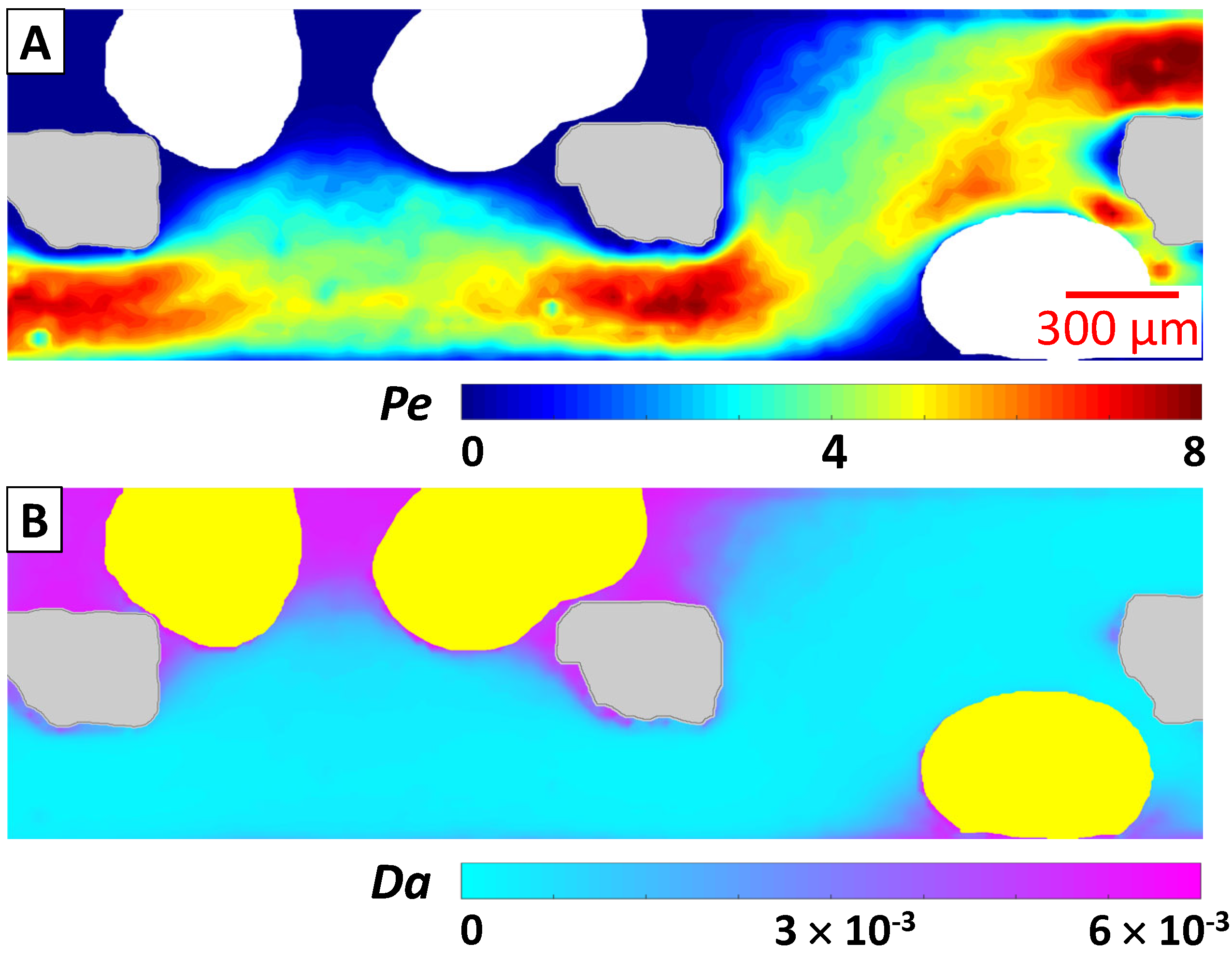}
\caption{$Pe$ (A) and $Da$ (B) fields calculated based on the velocity field in Fig.~\ref{fig:piv}B and Equations~\ref{eqn:Pe} and \ref{eqn:Da}.}
\label{fig:PeDa}
\end{figure}

\noindent Here, $V$ is the fluid velocity, $l$ is the characteristic length scale (\textit{e.g.}, pore diameter), $D$ is the diffusion coefficient, and $k$ is the reaction rate constant. Physically, $Pe$ defines the ratio of advective to diffusive transport rates, and $Da$ defines the ratio of the overall chemical reaction rate to the advective mass transport rate. Fig.~\ref{fig:PeDa} shows the $Pe$ and $Da$ fields calculated with Eqs. ~\ref{eqn:Pe} and ~\ref{eqn:Da} based on the velocity field shown in Fig.~\ref{fig:piv}B, highlighting the spatial variability of these fields, which will be valuable in understanding transport and reaction potential \cite{Jimenez2020, Menke2016}.

\begin{figure}[hbt]
\centering
\includegraphics[width=0.9\linewidth]{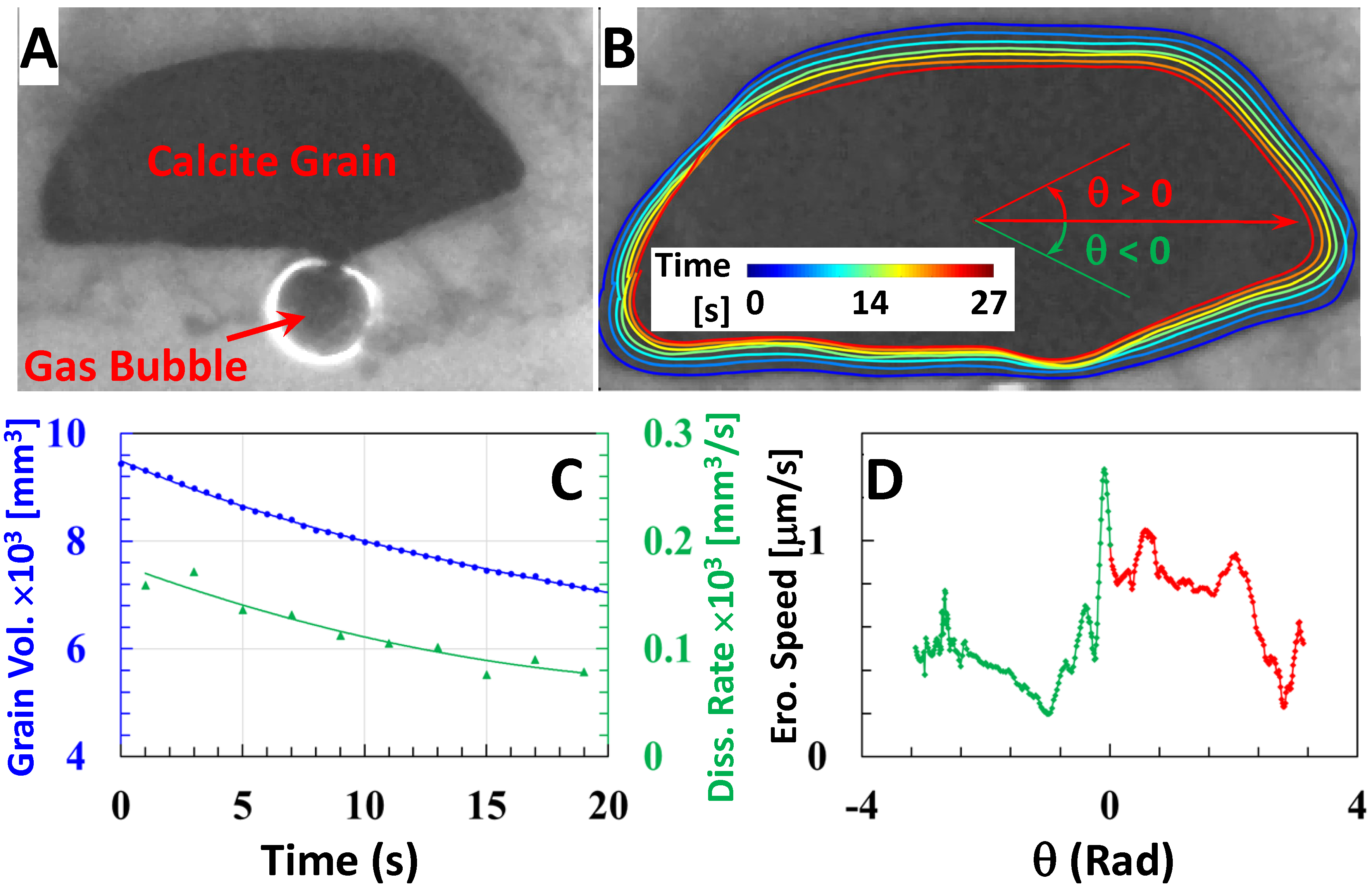}
\caption{Dissolution of a particular grain: (A) grain at $t = 0$\,s, (B) evolution of grain boundaries with time, (C) temporal variation of total grain volume and overall dissolution rate and (D) spatial variation of local dissolution rate around the grain.}
\label{fig:DissolutionRate}
\end{figure}

Another key attribute of the this flow is the pore-scale dissolution rate of grain. Fig.~\ref{fig:DissolutionRate} depicts the dissolution process of a particular grain subject to a flow of 0.8\% HCl solution at 0.2\,ml/min. Fig.~\ref{fig:DissolutionRate}A acquired with the white-field microscopy shows at an instance ($t = 0$\,s) the geometry of the grain with a CO$_2$ bubble attached to its bottom. Bubbles are observed to preferentially grow at this particular location, presumably due to the relatively sharp corner therein, which serves as nucleation site. Fig.~\ref{fig:DissolutionRate}B illustrates the evolution of the solid boundaries with time. These plots clearly indicate a much faster erosion of the grain at its front than its back due to stronger convective transport in the upstream. It also shows that the dissolution rate at the bottom of the grain corresponding to the location of the bubble in Fig.~\ref{fig:DissolutionRate}A is very low, suggesting that the presence of CO$_2$ significantly impeded dissolution therein and the neighboring region adjacent to it. Fig.~\ref{fig:DissolutionRate}C plots the temporal variation of the total grain volume and the overall dissolution rate as a function of time, whereas Fig.~\ref{fig:DissolutionRate}D plots the spatial variation of the dissolution rate around the grain at a certain moment ($t = 24$\,s). While both the total grain volume and the overall dissolution rate decrease with time in a quite smooth manner, the local dissolution rate displays a large variability owing to the complex flow fields and disturbance of CO$_2$ bubble around the grain. This behavior again reinforces the importance of the characterization of local dissolution and pore flow, which has never been achieved before.

\begin{figure}[hbt]
\centering
\includegraphics[width=1\linewidth]{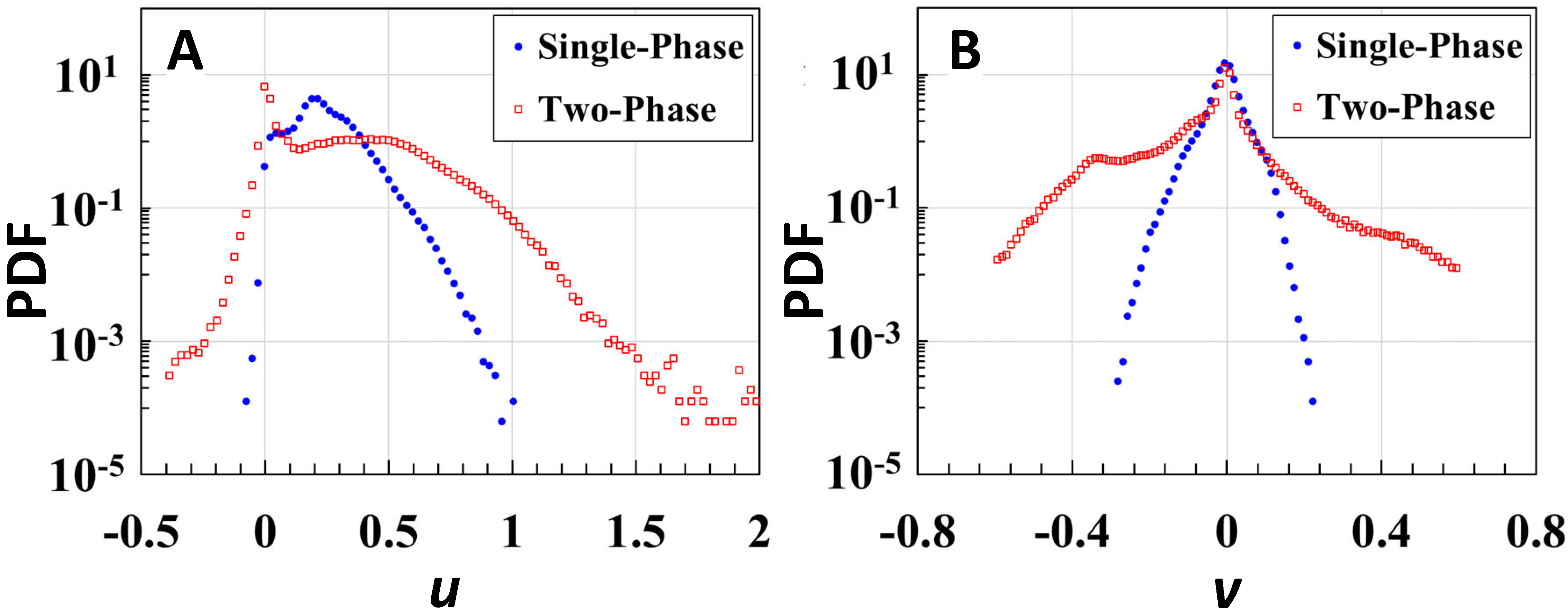}
\caption{PDFs: (A) horizontal component, $u$ and (B) vertical component, $v$ for the single and multiphase flows shown in Fig.~\ref{fig:piv}.}
\label{fig:PDF}
\end{figure}

The spatially- and temporally-resolved velocity data also enables the computation of flow statistics, providing insight into the integrated effects of pore-scale process thus can be particularly useful to develop constitutive correlations necessary for large-scale predictive models.  Fig.~\ref{fig:PDF} presents the probability density functions (PDF) of normalized velocity components, $u$ and $v$ in a single- and multiphase flow, respectively. The PDFs reveal that a second phase renders the flow much more random with broader distributions for both $u$ and $v$. These results also highlight the flow alteration induced by multiphase flow, justifying the need for systematic studies of the subtle effects of multiphase flow on mineral dissolution. In fact, statistical measures, specifically PDFs, have also been used as a great tool to characterize pore structures in order to reveal underlying physics that is not apparent \textit{via} local observations \cite{Jimenez2020, Al2017}. Additionally, PDFs also provide another convenient basis for experimental validation of computer codes for multiphase fluid flow. As pointed out by Meakin et al. \cite{Meakin2009}, due to fabrication error, microscale roughness, trace impurities and other experimental and numerical uncertainties, pore by pore comparison is virtually impractical, and thus comparison of statistical measures such as PDFs is often desired.

\section{Conclusions}
In this study, a calcite-based micromodel was successfully designed and fabricated using microfabrication techniques. The $\mu$PIV technique was employed that captured spatially and temporally resolved dynamics of single- and multiphase flows of CO$_2$ and diluted HCl in 2D calcite-based porous media. While the flow field with single-phase flow is relatively simple, that flow is significantly modified in the multiphase flow case due to the presence CO$_2$ bubbles that are generated in-situ as a result of chemical reaction between the liquid and solid phases. The separate CO$_2$ phase is expected to not only divert the HCl flow, but also shied the solid surfaces from further reaction, thus significantly modifying the local dissolution pattern and rate. The preliminary results not only provide a unique view of the flow dynamics during multiphase mineral dissolution, but also demonstrate great potential of this methodology for future more in-depth studies. 

\section*{Author Contributions}

\section*{Conflicts of interest}
There are no conflicts to declare.

\section*{Acknowledgements}
This work was performed in part at the Montana Nanotechnology Facility, an NNCI facility supported by NSF Grant ECCS-1542210, and with support by the Murdock Charitable Trust. The work was partially supported by the Petroleum Research Fund, ACS (62687-DNI9YL) with Dr. Thomas Clancy serving as the program officer. YL thanks the Norm Asbjornson College of Engineering and the Center for Faculty Excellence at Montana State University for their support through the Faculty Excellence Grants. RMR thanks the Norm Asbjornson College of Engineering for their support through the Benjamin Fellowship.





\bibliography{References.bib} 
\bibliographystyle{rsc} 

\end{document}